\documentclass[11pt]{article}

\usepackage{xspace}
\usepackage{color}
\usepackage{url}
\usepackage{comment}

\newcommand{\ais}{ML/AI system\xspace}
\newcommand{\gais}[1]{generative ML/AI system#1\xspace}
\newcommand{\Gais}[1]{Generative ML/AI system#1\xspace}

\title{Copyright related risks in the creation and use of \ais{s}}
\author{Daniel M. German}

\begin{document}

\maketitle

\section{Introduction}

Without training data, a ML/AI system is useless.  What data, where this data comes from, and who owns this data are
some of the most important questions that should be answered before an ML/AI system is created. In some cases, this data
might be owned by the creator of the \ais (such as records of interaction with customers, historical records of
transactions, data explicitly created to train such system, etc).  In others, the data might by own by a another party
and therefore, this might impose restrictions on the creation and operation of the \ais.

\Gais{s} create works, and many of these works are expected to be used. Whether such works have an owner, and who
this owner is, are also important consideration for those who use \gais{s}.

Intellectual property (IP) legislation protects \emph{creations of the mind}, i.e. non-tangible property, such as
training data, \ais{s} and potentially the output these systems create.  IP is divided into several types, primarily
four: copyright, trademarks, patents and trade-secrets\cite{WIPOwhatIsIP}. These four types of IP can affect 
\ais{s}, but  due to space restrictions, this article concentrates solely on copyright.

IP legislation is national in scope, and thus, varies from country to country. Through the World Intellectual
Property Organization (WIPO)---a United Nations agency---its 193 member countries try to homogenize their IP
legislations. Nonetheless, there exist major differences between countries IP protections that
concern the creation, training and use of \ais{s}.

As of today, there exist significant legal uncertainty regarding the use of IP in the creation and use of an \ais.  In
December 2019, WIPO started a consultation process regarding AI and IP policy inviting member countries IP offices,
individuals and organizations to submit comments to 13 different issues\cite{wipoIssuesIPAI,
  WipoAIIPconsulationSubmisions}. Its long term goal is to publish a series of documents that can be used as guidelines
for its of its member countries.

In the meantime, current IP laws are being interpreted by the different \ais's stakeholders, their corresponding
countries' IP offices and courts. Several
lawsuits are in progress in various jurisdictions; their outcome is likely to affect the development and
use of \ais{s} (at least until, and if, IP legislation is changed to adapt to these new challenges).

This article describes the potential security risks (regarding availability) around the creation and use of \ais{s}. It
is divided into three sections: first, the concerns of the owners of copyrighted works used in training; second, the concerns of
the those who use generative \ais{s} to incorporate its output into copyrightable expression; ending with a set of
recommendations to help mitigate these risks.

\section{Concerns of owners of copyrighted works}

To understand how and under which conditions copyrighted material can be used for training, 
three fundamental questions need to be answered:

\begin{enumerate}
\item Does the use of copyrighted works in the training of \ais{s} require permission from its owners?
\item Is the trained AI/ML model a derivative work of the works it uses for training?
\item Are the creations of the \ais derivative works of the works used for training?
\end{enumerate}

Copyright laws across the world vary in how they restrict the use of copyrighted work without permission from
its owner. One notable example is the United States, where the doctrine of fair use determines whether a given use of
copyrighted work is permissible.  Fair use hinges on four key conditions: the purpose of the use, the nature of the
copyrighted work, the amount and substance used, and the effect on the potential market of the work. In contrast, other
countries define the concept of fair dealing, which is based on the reasons why the copy is made. For example, in the UK
and Canada fair-dealing is allowed for purposes such as research, private study, criticism, review, and news
reporting. Current copyright laws \emph{might allow} the use of copyrighted works in the training of \ais{s} as
long as their use is \emph{fair}.

Japan is an exception on this regard. In 2018, Japan's copyright law was modified to permit the use of copyrighted works
without permission for the purpose of information analysis---with few limitations \cite[Section 30-4]{japanCopyrightLaw}. This
permission allows the use of
copyrighted works without a license for training of models. As expected, copyright owners have raised concerns, and the government is
evaluating potential changes to this legislation\cite{JapanNewsAIPolicy23}.

Fair use legislation  leaves it to the courts to decide if a use if fair. It is therefore not surprising that several
lawsuits addressing this issue are already in progress.  For example, in Anderson v. Stability AI
Ltd\cite{andersenVsStability22}, the plaintiffs argue that the use of copyrighted images in the training of Stable
Fusion and Midjourney is a violation of their copyright, and that the images created by these AI systems are derivative
works of their copyrighted images. The plaintiffs request that these \ais{s} stop using their copyrighted
works and award them damages.

Getty Images has also objected to the use of its works in Stable Fusion. In its lawsuit \cite{gettyVsStability23}, Getty argues that the training of 
Stable Fusion has used its images, their captions and metadata without a license. An important argument that Getty makes
is that Stability AI (the owner of Stable Fusion) is competing with Getty in the same market (providing images to third
parties) thus weakening a potential defence based on fair use.

In another case, DOE 1 et al v. GitHub, Inc. et al\cite{doeVsGithub22}, the plaintiffs argue that both Github Copilot and OpenAI Codex have
violated the license of the open source programs they have used for training.

Thus, three main risks for copyright owners arise:

\begin{enumerate}
\item Copyright owners consider that \ais{s} use (and potentially incorporate) their works in an unfair manner.
\item Copyright owners perceive a potential loss of their market when AI generated content can replace theirs.
\item Copyright owners argue that it in some cases the generated works are derivative works of theirs.
\end{enumerate}

Note that some of the issues also affect those that publish trained models for others to use (such as those published in
Kaggle\footnote{\url{http://Kaggle.com}}). Frequently these models
come with a license that covers how the model can be used, but this license usually ignores  that the model might have
used copyrighted works for its training.

Owners of the copyright of works used for training have also objected to the way their works are harvested, claiming
that this is done in breach of contracts. For example, in its lawsuit against
Stability AI, Getty Images claims that the
LAION public datasets (sponsored by Stability AI) contain ``12 million links to images their associated text and
metadata'' and this has information has been used by Stability AI to copy high resolution images and their detailed
metadata breaking the terms and conditions of its website (which explicitly prohibit these types of data harvesting)\cite{gettyVsStability23}.

Collections of copyrightable and non-copyrightable works can also be protected by copyright.
In the UK and the European Union this protection grant explicit copyright protection to databases\cite{ukCopyrightLaw,dbDirectiveEU}. In the United States
similar works are protected as compilations. Database/compilation protection might
be claimed by owners datasets used during training.

Consequently, the risks for ML/AI systems creators and maintainers are:

\begin{enumerate}
\item Copyright owners (including owners of databases) might claim a violation of their copyright on works use for the training of the system.
\item Web site operators might claim a breach of contract by those who gather copyrighted works for training.
\item It might be up to a court of law to decided if the use of the works is fair, thus necessitating a trial.
\item Using a pre-trained model might have an associated legal risk if the training was done using copyrighted
  works without the permission from its owners.
\end{enumerate}

\section{Concerns of users who incorporate content generated by \ais{s} into their creations}

Many users are already incorporating generated content into their expression. From their point of view, two fundamental
questions need to be answered:

\begin{enumerate}
\item  Who is the owner of the generated work?
\item Is the generated work copyrightable? 
\end{enumerate}

As mentioned above, in DOE 1 et al v. GitHub, Inc. et al \cite{doeVsGithub22} the owners of the copyright of source
code used for training believe that some generated works are derivative works of theirs, and thus, are subject to the
terms of the license of their software (in this particular case mostly an open source
license) and such license might have an impact on the licenses that the generated work can be licensed
under (e.g. if a given source code used in training is licensed under the General Public License v 2.0, and the generated
code includes portions of this source code, then the generated code should also be
licensed under the same license). Microsoft (owner of Github), recognizing that this perceived risk might affect the
market of Copilot recently issued the \emph{Copilot Copyright Commitment}: ``Microsoft’s existing IP indemnification coverage to
copyright claims relating to the use of our AI-powered Copilots, including the output they generate, specifically for
paid versions of Microsoft commercial Copilot services and Bing Chat Enterprise.'' \cite{microsoftCopilotComm23}.

With respect to the copyrightability of generated works, it is important to remember that copyright is granted at the moment a
work is created (``fixed in a tangible medium'' as long as it is ``an original work of authorship'' \cite{usCopyrightLaw}). Thus, the question
of whether the generated work of ML/AI system is copyrightable should be rephrased as: \emph{Would courts of law reject the claim that the generated work has a copyright?}

In the US, before a lawsuit is filled on behalf of a copyright owner, the owner must obtain registration of the work in
the Copyright Office.  The US Copyright Office would evaluate the claim and determine if the work deserves
copyright. Recently, the Copyright Office issued a set of guidelines  stating the requirements for registration of works created with the assistance of 
AI\cite{usCopyRegis23}. In a nutshell, that copyright cannot be granted to the output of a \ais, but that copyright 
can be granted to works created with the assistance of an \ais as long as there is sufficient creativity in combining
the generated and non-generated content into a new work. The copyright office requires, in the application for copyright
registration, that the creator exclude from the application ``any non-insignificant portions generated [by the AI/ML
system]'' \cite{usCopyRegis23}. It also clarifies that the owner of the copyright of the \ais cannot claim copyright on
the work being registered.

A famous work created with the help of two AI/ML systems called \emph{Théâtre D’opéra Spatial} won the 2022 Colorado State Fair
Annual Art Competition under the category ``digital arts/digitally manipulated photograph''\cite{theatreOpera}. Its creator filled for
US copyright registration without disclosing that the
work had been created with the help of ML/AI systems. The Copyright Office, due to the media attention received by the
work, was aware that
it had been created with the help of two ML/AI systems and requested clarification from its author 
regarding the contributions of the \ais{s} to the overall work.  It responds reads:
``Upon analysis, the Copyright Office decided that, based on the application for copyright, the work does not deserve,
copyright protection because "the Work" contains more than a de minimis amount of content generated by artificial
intelligence (“AI”), and this content must therefore be disclaimed in the application for registration. Because
[the author] is unwilling to disclaim the AI-generated material, the Work cannot be registered as submitted.'' 

In another case, the US Copyright office accepted the copyright registration of a graphic novel but rejected the
registration of the individual images as deserving copyright because ``the images in the Work that were generated by the
Midjourney technology are not the product of human authorship.'' \cite{zaryaUScopy23}. Without copyright, a work is
effectively in the public domain.

The US Copyright Office has clarified that the prompts used to generate the work ``function more like instructions to a
commissioned artist—they identify what the prompter wishes to have depicted, but the machine determines how those
instructions are implemented in its output.''. Therefore, the creativity (and copyright) of prompts cannot be used to claim
copyright of its corresponding generated work.

The United Kingdom has taken an opposite view. Copyright is granted to computer generated literary, dramatic, musical or artistic
works: ``the author shall be taken to be the person by whom the arrangements necessary for the creation of the
work are undertaken.'' \cite{ukCopyrightLaw}. The UK Intellectual Property Office has been having a
consultation regarding AI and current IP laws, and while it acknowledges that there exist many unresolved issues, it
stated that there are no plans to modify the law regarding computer generated works \cite{ukConsultationOutcome22}.

The operators of \ais{s} might also claim copyright on the created works.
Midjourney terms of use indicate that Midjourney Inc. owns the copyright of generated works when the user does not pay
for the service \cite{midjourneyTerms23}, and these works can only be used for non-commercial purposes.

Thus, the risks for a creator of works using an ML/AI systems are:

\begin{enumerate}
\item The owners of the training data (and where applicable, the owners of the trained model) might claim derivative
  copyright of the generated work.
\item The operator of the \ais  might claim ownership of the generated work.
\item The creator might not be able to gain copyright on the generate work.  
\item The interaction with the \ais (the prompts) is likely to be protected by copyright, but these prompts might have no impact on the
    copyright of the generated work.
\item Even if copyright is originally granted, a review of the copyright registration might lead towards its loss.
\end{enumerate}

\section{Mitigating the risks}

Across the world, there is significant legal uncertainty that affects the owners of training data, and the operators and
users of \ais{s}. And these risks vary from jurisdiction to jurisdiction. Stakeholders must familiarize with their
corresponding legislations.

Owners of training data can take advantage of other laws to increase the protection of their creations, such as trademarks, contracts
that limit the copying of their works, and trade secret legislation. For example, 
in Getty v. Stability Inc\cite{gettyVsStability23}, Getty argues for trademark dilution (among other claims, as
described above).

Publishers of trained models, and operators of \ais{s} could be liable for contributory infringement, and thus, should
consider carefully the origin and ownership of the data used for training.

Finally, the users of generative \ais{s} are potentially the ones at the biggest risk. Depending on the jurisdiction where
they operate, they should carefully consider the limitations imposed by the use of generated works into their
creations. As exemplified above, in the United States it is not currently possible to gain copyright for works, or portions of
works, created by a \ais, and therefore, the copyright of a work will not include portions that were generated. In some
domains (such as visual art) this might be a major limitation.

Trademarks and contracts (e.g. terms of use) likely have an impact on the use of \ais{s} and the works they generate, and therefore,
users should familiarize with their terms of use. These terms might impose certain restrictions and conditions regarding
ownership of generated works, For instance, ChatGPT terms of use clarify that the user owns the generated works:
``Subject to your compliance with these Terms, OpenAI hereby assigns to you all its right, title and interest in and to
Output.'' However, it also warns that the same output might be generated by different
users. \cite{chatGPTterms23}. Midjourney takes an opposite approach that covers the prompts and the
generated work: ``You grant to Midjourney [... a] copyright license to [...] \textbf{text}, and \textbf{image prompts You input} into the
Services, or \textbf{Assets produced} by the service at Your direction''  and ``If You are not a Paid Member, \textbf{You
don’t own the Assets You create}. Instead Midjourney grants you a [...]  Creative Commons \textbf{Noncommercial} 4.0 Attribution International License'' (emphasis added)  \cite{midjourneyTerms23}.

Countries might adapt their copyright and other IP-related laws to balance the economic rights of the owners of the training
data with the potential benefits of \ais{s}. At the very least they will continue to publish guidelines that address
common issues. In the meantime lawsuits (and to a lesser extend contracts) will continue to guide the perception of what is
acceptable use of copyrightable works in the training of \ais{s}, and what generated output is copyrightable, and who
its owner is (even if these lawsuits are settled
out of court). \ais{s} operators and their users must carefully evaluate and manage these risks.

``AI is a fast-evolving technology with great potential to make workers more productive, to make
firms more efficient and to spur innovations in new products and services''\cite{aiImpactEuUS}. The courts and legislators will have a big
influence on whether and how this potential is achieved.

\bibliography{references}

\begin{thebibliography}{10}

\bibitem{WIPOwhatIsIP}
World Intellectual~Property Organization.
\newblock What is intellectual property, 2004.
\newblock WIPO Publication No. 450E/20.

\bibitem{wipoIssuesIPAI}
World Intellectual Property~Organization (WIPO).
\newblock Draft issues paper on intellectual property policy and artificial
  intelligence.
\newblock WIPO/IP/AI/2/GE/20/1, Dec 2019.
\newblock Main issues of IP and AI.

\bibitem{WipoAIIPconsulationSubmisions}
World Intellectual~Property Organization.
\newblock {The WIPO Conversation on Intellectual Property and Artificial
  Intelligence}.
\newblock
  \url{https://www.wipo.int/about-ip/en/artificial_intelligence/conversation.html}.
\newblock Accessed Sept 2024.

\bibitem{japanCopyrightLaw}
Japan~National Diet.
\newblock {Copyright Act}, 2020.

\bibitem{JapanNewsAIPolicy23}
The~Yomieru Shinbun.
\newblock {Intellectual Property Plan Signals Reversal on AI Policy}.
\newblock Japan News, June 10, 2023.

\bibitem{andersenVsStability22}
{Andersen v. Stability AI Ltd.}
\newblock Andersen v. Stability AI Ltd., 3:23-cv-00201, (N.D. Cal.).

\bibitem{gettyVsStability23}
{Getty Images (US) Inc v Stability AI Inc}.
\newblock 1:23-cv-00135, (D. Del.).

\bibitem{doeVsGithub22}
{DOE 1 et al v. GitHub}.
\newblock 4:22-cv-06823, (N.D. Cal.).

\bibitem{ukCopyrightLaw}
Parliament:~House of~Commons.
\newblock {Copyright, Designs and Patents Act 1988 chapter 48}, 1988.

\bibitem{dbDirectiveEU}
{The European Parliament and the Council of the European Union}.
\newblock Directive 96/9/ec, March 1996.

\bibitem{microsoftCopilotComm23}
Brad Smith and Hossein Nowbar.
\newblock {Microsoft announces new Copilot Copyright Commitment for customers}.
\newblock Microsoft On the Issues Blob,
  \url{https://blogs.microsoft.com/on-the-issues/2023/09/07/copilot-copyright-commitment-ai-legal-concerns/}.

\bibitem{usCopyrightLaw}
U.S. Congress.
\newblock {United States Code, 2018 Edition, Supplement 3, Title 17 -
  Copyrights}, 2021.

\bibitem{usCopyRegis23}
United States~Copyright Office.
\newblock Copyright registration guidance: Works containing material generated
  by artificial intelligence, March 2023.
\newblock 16190 Federal Register, Vol. 88, No. 51.

\bibitem{theatreOpera}
Wikipedia.
\newblock {Théâtre d'Opéra Spatial}.
\newblock \url{https://en.wikipedia.org/wiki/Théâtre_d%27Opéra_Spatial}.

\bibitem{zaryaUScopy23}
United States~Copyright Office.
\newblock Re: Zarya of the dawn (registration \# vau001480196), Feb 2023.

\bibitem{ukConsultationOutcome22}
Intellectual~Property Office.
\newblock {Artificial Intelligence and Intellectual Property: copyright and
  patents: Government response to consultation}, June 2022.

\bibitem{midjourneyTerms23}
Midjourney Inc.
\newblock {Midjourney Terms of Use}.
\newblock \url{https://docs.midjourney.com/docs/terms-of-service} Accessed
  Sept. 2023.

\bibitem{chatGPTterms23}
OpenAI.
\newblock {OpenAI Terms of Use}.
\newblock \url{https://openai.com/policies/terms-of-use} Accessed Sept. 2023.

\bibitem{aiImpactEuUS}
US-EU Trade and Technology Council.
\newblock {The Impact of Artificial Intelligence on the Future of Workforces in
  the European Union and the United States Of America}, Dec. 5, 2022.

\end{thebibliography}

\bibliographystyle{unsrt}

\end{document}